\documentclass[journal]{IEEEtran}
\usepackage{graphicx,amsmath,subfigure,epstopdf,color}
\usepackage{fancyhdr}
\renewcommand{\footskip}{20pt}
\begin{document}
\title{Integral Equations for Computing AC Losses of Radially and Polygonally Arranged HTS Thin Tapes}
%
%
% author names and IEEE memberships
% note positions of commas and nonbreaking spaces ( ~ ) LaTeX will not break
% a structure at a ~ so this keeps an author's name from being broken across
% two lines.
% use \thanks{} to gain access to the first footnote area
% a separate \thanks must be used for each paragraph as LaTeX2e's \thanks
% was not built to handle multiple paragraphs
%

\author{Roberto Brambilla, Francesco Grilli, and Luciano Martini% <-this % stops a space
\thanks{R. Brambilla and L. Martini are with RSE -- Ricerca sul Sistema Energetico S.p.A., Milano, Italy (e-mail: roberto.brambilla@rse-web.it, luciano.martini@rse-web.it). F. Grilli is with Karlsruhe Institute of Technology, Karlsruhe, Germany (e-mail: francesco.grilli@kit.edu).}
\thanks{This work has been supported partly by the Research Fund for the Italian Electrical System under the Contract Agreement between RSE and the Ministry of Economic Development - General Directorate for Nuclear Energy, Renewable Energy and Energy Efficiency stipulated on July 29, 2009 in compliance with the Decree of March 19, 2009, and partly by the Helmholtz-University Young Investigator Grant VH-NG-617.}% <-this % stops a space
}

\maketitle

\begin{abstract}
%\boldmath
In this paper we derive the integral equations for radially and polygonally arranged  high-temperature superconductor thin tapes and we solve them by finite-element method. The superconductor is modeled with a non-linear power law, which allows the possibility of considering the dependence of the parameters on the magnetic field or the position. The ac losses are computed for a variety of geometrical configurations and for various values of the transport current. Differences with respect to existing analytical models, which are developed in the framework of the critical state model and only for certain values of the transport current, are pointed out.
\end{abstract}
% IEEEtran.cls defaults to using nonbold math in the Abstract.
% This preserves the distinction between vectors and scalars. However,
% if the journal you are submitting to favors bold math in the abstract,
% then you can use LaTeX's standard command \boldmath at the very start
% of the abstract to achieve this. Many IEEE journals frown on math
% in the abstract anyway.

% Note that keywords are not normally used for peerreview papers.

% For peer review papers, you can put extra information on the cover
% page as needed:
% \ifCLASSOPTIONpeerreview
% \begin{center} \bfseries EDICS Category: 3-BBND \end{center}
% \fi
%
% For peerreview papers, this IEEEtran command inserts a page break and
% creates the second title. It will be ignored for other modes.
\IEEEpeerreviewmaketitle

\section{Introduction}\label{sec:introduction}
% The very first letter is a 2 line initial drop letter followed
% by the rest of the first word in caps.
% 
% form to use if the first word consists of a single letter:
% \IEEEPARstart{A}{demo} file is ....
% 
% form to use if you need the single drop letter followed by
% normal text (unknown if ever used by IEEE):
% \IEEEPARstart{A}{}demo file is ....
% 
% Some journals put the first two words in caps:
% \IEEEPARstart{T}{his demo} file is ....
% 
% Here we have the typical use of a "T" for an initial drop letter
% and "HIS" in caps to complete the first word.
\IEEEPARstart{T}{he} current density distribution and the ac losses of thin superconductors carrying ac current and/or subjected to external ac magnetic field can be  calculated by solving the integral equations for the current density distribution by means of finite elements~\cite{Brambilla:SST08}. The integral equation (IE) model was first proposed in~\cite{Brambilla:SST08} for individual tapes; then, it was extended to the case of interacting tapes in~\cite{Grilli:TAS09a}, where the electromagnetic interaction between tapes was calculated by means of an auxiliary 2-D magnetostatic model; finally, it was developed to the stage where the interaction between tapes is directly included in the integral equation to be solved~\cite{Brambilla:SST09}.

The most important advantages of this approach, especially compared to analytical formulations, are that arbitrary current-voltage characteristics for the superconductor and the dependence of the superconductors parameters (e.g. the critical current density $J_c$) on the position and the local magnetic field can be easily incorporated.

The IE model has been successfully used to compute the ac losses of interacting tapes and a very good agreement with experimental data has been found for tapes characterized by a lateral variation of $J_c$~\cite{Nguyen:SST09} and arranged in a bifilar winding for fault-current limiter application~\cite{Elschner:PhysC12}.

In this paper we use the approach utilized in~\cite{Brambilla:SST09} to derive and solve the integral equations for the current density distribution in two other cases of practical interest, namely cables with radially and polygonally arranged thin superconducting tapes. The former is potentially interesting for bus bar applications (especially with bidirectional currents, see~\cite{Kato:PISS93, Ando:TAS95}); the latter can be useful to quickly estimate the losses of straight cable samples~\cite{Amemiya:TAS07a}. These configurations were analyzed by Mawatari and Kajikawa under the assumptions of the critical state model~\cite{Mawatari:APL06, Mawatari:APL08}. Their  method is not devoid of elegance, however their solutions are valid only for certain current levels ($I \ll I_c$ for both configurations and $I=I_c$ for the polygonal configuration only), which are usually not found in practical applications. On the contrary, the IE approach allows using arbitrary current levels. In addition, it can take the dependence of certain parameters on the local magnetic field and/or on the position inside the tape into account, see for example~\cite{Nguyen:SST09} and~\cite{Grilli:SST10a}. This latter dependence is very important for simulating real ReBCO coated conductors, which often present a reduced $J_c$ near the edges due to the manufacturing process~\cite{Amemiya:PhysC06b, Haenisch:SST08}.

The paper is organized as follows. Section~\ref{sec:derivation} describes in detail the derivation of the integral equations for the cases under scrutiny. Section~\ref{sec:results} contains the main results of our investigation: we discuss the dependence of the ac losses on the geometrical parameters and we show the obtained current profiles; differences from and analogies with the analytical models are pointed out. Finally, section~\ref{sec:conclusion} draws the main conclusions.

\section{Derivation of the integral equations}\label{sec:derivation}
\begin{figure}[t]
\centering
\includegraphics[width=0.7\columnwidth]{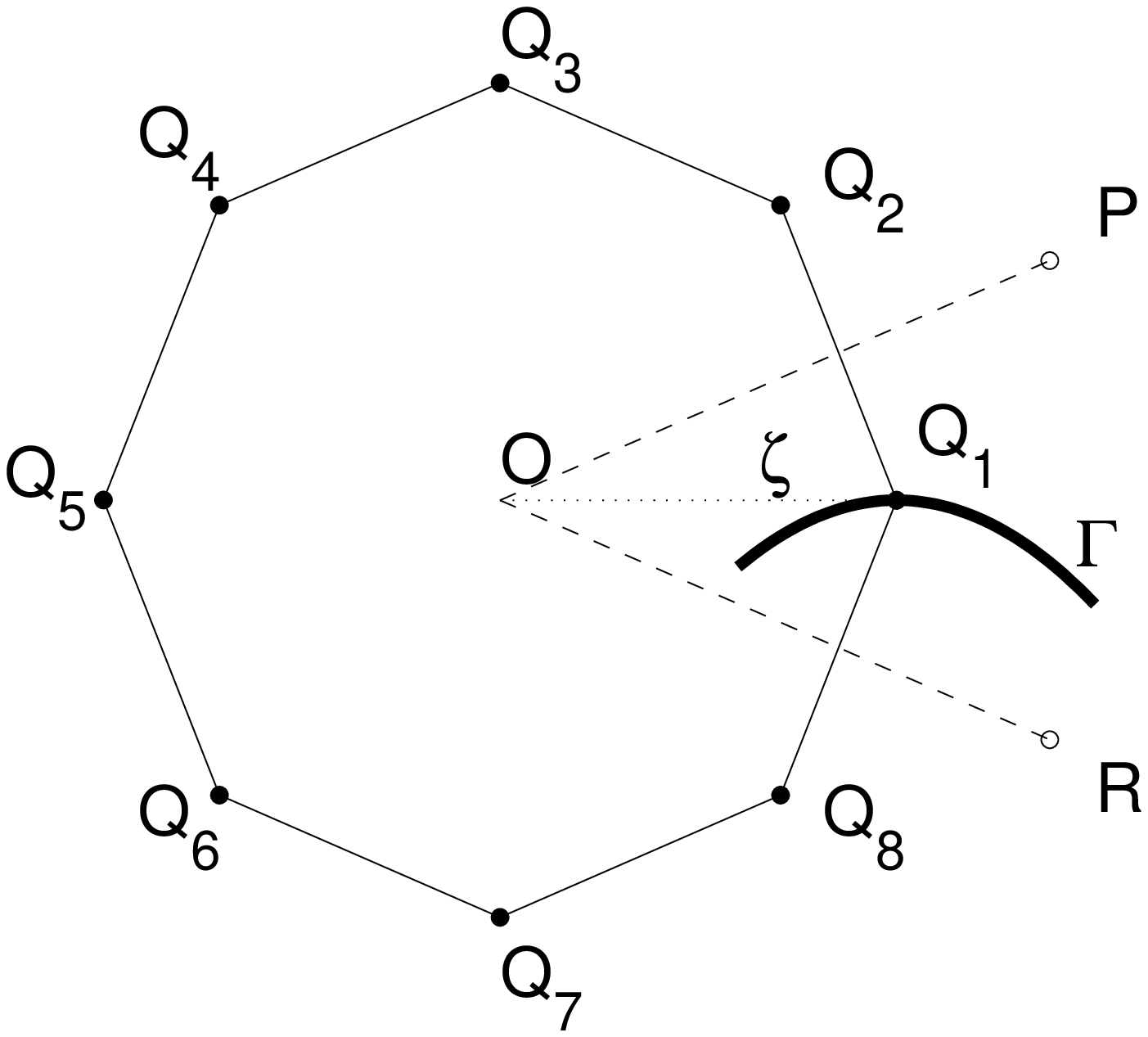}
\caption{Reference periodical geometry used for the derivation of the integral equations -- see section~\ref{sec:derivation}.}
\label{fig:periodicity}
\end{figure}
In the complex plane, we consider the vertices $Q_i$ of a regular polygon with $n$ sides centered in the origin $O$ and a point $P$ as shown in Fig.~\ref{fig:periodicity}. If we call $z$ the point $P$ and $\zeta$ the point $Q_1$, the vertices of the polygon will be given by $Q_k=\omega_{k-1}\zeta$, where $\omega_k=\exp (2 \pi i k/n)$ are the $n$-th roots of the unity. Based on the properties of those roots, we shall have
\begin{equation}
\prod_{k=0}^{n-1}(z-\omega_k \zeta)=z^n-\zeta^n.
\end{equation}
Taking the logarithm of both members, we  have
\begin{equation}
\sum_{k=0}^{n-1}\log(z-\omega_k \zeta)=\log(z^n-\zeta^n)
\end{equation}
and by deriving with respect to $z$ we obtain the identity
\begin{equation}\label{eq:sum}
\sum_{k=0}^{n-1}\frac{1}{z-\omega_k \zeta}=\frac{n z^{n-1}}{z^n-\zeta^n}.
\end{equation}
\subsection{Magnetic field generated by current lines arranged with angular periodicity}
Since in two-dimensions the Biot-Savart law for the magnetic field generated by a current line $I_0$ situated in $\zeta$ can be expressed in the complex plane by the following complex expression
\begin{equation}
H(z)=H_y(z)+iH_x(z)=\frac{I_0}{2\pi}\frac{1}{z-\zeta},
\end{equation}
we shall have that, according to (\ref{eq:sum}), the magnetic field generated by $n$ {\it identical} current lines situated in the vertices $\omega_k \zeta$ of a regular polygon of radius $\zeta$ is
\begin{equation}\label{eq:Hzn}
H(z) = \frac{I_0}{2 \pi}\sum_{k=0}^{n-1}\frac{1}{z-\omega_k \zeta}=\frac{I_0}{2 \pi}\frac{n z^{n-1}}{z^n-\zeta^n}.
\end{equation}
In the case where $n=2m$ and the currents have alternate signs, we can write
\begin{align}
\nonumber
H(z) &= \frac{I_0}{2 \pi}\sum_{k=0}^{n-1}\frac{(-1)^k}{z-\omega_k \zeta}  \\ 
\nonumber
&= \frac{I_0}{2 \pi}\left ( \sum_{k=0}^{m-1}\frac{1}{z-\omega_{2k}\zeta}  - \sum_{k=0}^{m-1}\frac{1}{z-\omega_{2k+1}\zeta} \right ) \\
&=\frac{I_0}{2 \pi}\left ( \sum_{k=0}^{m-1}\frac{1}{z-\omega_{2k}\zeta}  - \sum_{k=0}^{m-1}\frac{1}{z-\omega_{2k}(\omega_1 \zeta)} \right ).
\end{align}
Using (\ref{eq:sum}) and the fact that $\omega_1^m=\omega_1^{n/2}=\exp(i\pi)=-1$, for $m$ currents we can write
\begin{align}\label{eq:Hzm}
\nonumber
H(z) &= \frac{I_0}{2 \pi} \left ( \frac{mz^{m-1}}{z^m-\zeta^m}-\frac{mz^{m-1}}{z^m-\omega_1^m\zeta^m} \right ) \\
&=  \frac{I_0}{2 \pi} \left ( \frac{mz^{m-1}}{z^m-\zeta^m}-\frac{mz^{m-1}}{z^m+\zeta^m} \right ) = \frac{I_0}{2 \pi} \frac{nz^{n/2-1}\zeta^{n/2}}{z^n-\zeta^n}.
\end{align}

\subsection{Magnetic field generated by current line distributions arranged with angular periodicity}
Let us consider the case where the magnetic field is generated by a distribution of current lines $J(z)$ along a path $\Gamma$ with angular periodicity, as shown in Fig.~\ref{fig:periodicity}. The magnetic field is obtained simply by integrating (\ref{eq:Hzn}) or (\ref{eq:Hzm}) along $\Gamma$
\begin{equation}\label{eq:Hz_Gamma}
H(z) = \frac{1}{{2\pi }}\int\limits_\Gamma  {J(\zeta )\frac{{n\,{z^{n - 1}}}}{{{z^n} - {\zeta ^n}}}\,d\zeta }.
\end{equation}
Due to the angular periodicity it is sufficient to study the magnetic field in the angular sector $\alpha=2 \pi /n$.
We will now study two cases interesting for practical applications: a system of $n$ tapes with radial and polygonal arrangement, which are schematically represented in Fig.~\ref{fig:star-polygonal}.

\subsection{System of $n$ radially arranged tapes}
\begin{figure}[t!]
\centering
\subfigure[]
{
\label{fig:star}
\includegraphics[width=0.7\columnwidth]{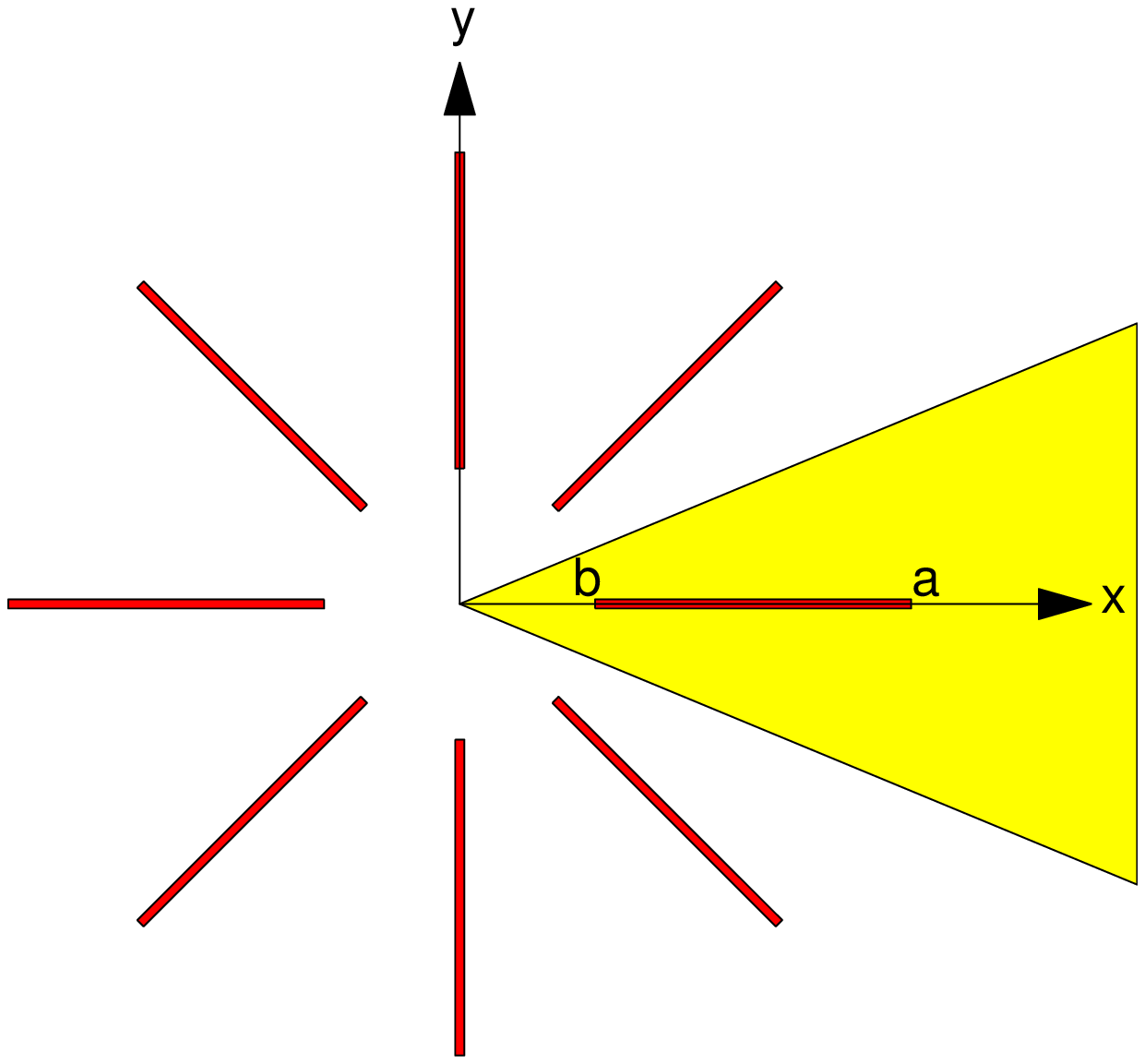}
}
\subfigure[]
{
\label{fig:polygonal}
\includegraphics[width=0.7\columnwidth]{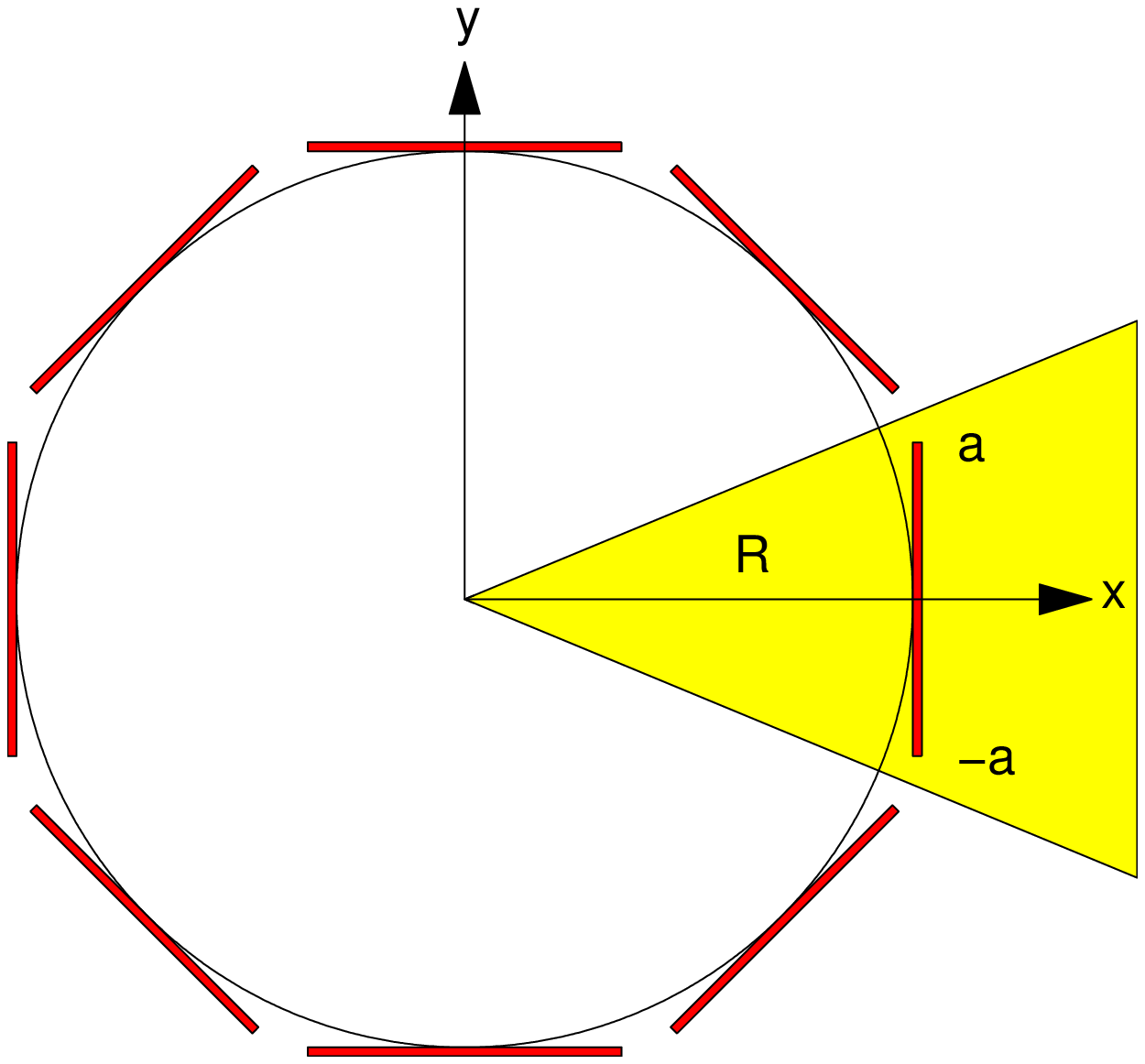}
}
\caption{Schematic drawings of the geometries considered in this work: (a) Radially-arranged and (b) polygonally-arranged superconducting thin tapes. For the radial arrangement, both unidirectional and bidirectional currents are considered.}
\label{fig:star-polygonal}
\end{figure}

Let us place the tape along a segment $(b,a)$ on the $x$-axis with a current density distribution $J(x)$ -- see Fig.~\ref{fig:star}. The magnetic field will be given by the integral
\begin{equation}
H(x) = \frac{1}{{2\pi }}\int\limits_b^a {J(\xi )\frac{{n\,{x^{n - 1}}}}{{{x^n} - {\xi ^n}}}\,d\xi }.
\end{equation}

We can compute the flux (in each tape) between the edge $b$ and an arbitrary point $x<a$ as	 
\begin{align}\label{eq:flux}
\nonumber
\Phi (x) &= \int\limits_b^x {{H_y}(t)dt}  = {\mathop{\rm Re}\nolimits} \int\limits_b^x {H(t)dt}  \\
\nonumber
&= \frac{1}{{2\pi }}\int\limits_b^x {dt} \int\limits_b^a {J(\xi )\frac{{n\,{t^{n - 1}}}}{{{t^n} - {\xi ^n}}}\,d\xi } \\
\nonumber
&= \frac{1}{{2\pi }}\int\limits_b^a {J(\xi )\,d\xi \int\limits_b^x {\frac{{n\,{t^{n - 1}}}}{{{t^n} - {\xi ^n}}}\,} dt}  \\
&= \frac{1}{{2\pi }}\int\limits_b^a {J(\xi )} \,\ln \left| {\frac{{{x^n} - {\xi ^n}}}{{{b^n} - {\xi ^n}}}} \right|\,d\xi.
\end{align}

Similarly to what was done in~\cite{Brambilla:SST08}, using this expression for the flux we can define the integral equation for the computation of the current density
\begin{equation}\label{eq:IE_unidir}
\rho J(x,t) = \mu d\frac{1}{{2\pi }}\int\limits_{ - a}^a {\dot J(\xi ,t)\ln \left| {{x^n} - {\xi ^n}} \right|} \,d\xi +C(t),
\end{equation}
where the term $C(t)$ is the contribution to the integral in~(\ref{eq:flux}) due to the
logarithmic part not dependent on the flux variable $x$. In practice the
$C(t)$ term is obtained by imposing the total current $I(t)$ flowing in each tape. For the solution of the integral equation by finite elements the term does not need to be explicitly introduced in the equation, since it is automatically evaluated by means of the integral constraint.

\begin{equation}\label{eq:IE_constraint}
\int\limits_{ - a}^a J(x,t) \,dx=I(t).
\end{equation}

In case of bidirectional currents (neighboring tapes carrying current of the same amplitude but with opposite direction), one can follow the same procedure to find the integral equation for the current density
\begin{equation}\label{eq:IE_bidir}
\rho J(x,t) = \mu d\frac{1}{{2\pi }}\int\limits_{ - a}^a {\dot J(\xi ,t)\ln \left| {\frac{{{x^m} - {\xi ^m}}}{{{x^m} + {\xi ^m}}}} \right|} \,d\xi+C(t).
\end{equation}
\subsection{System of $n$ polygonally arranged tapes}
In this case the tape can be represented by the segment $(-a,a)$ at a distance $R$ from the origin -- see Fig.~\ref{fig:polygonal}. Obviously, in order to avoid tape overlapping, the condition $R>a \cot(\pi/n)$ must hold. According to (\ref{eq:Hz_Gamma}) the magnetic field is given by the integral
\begin{equation}
H(z) = \frac{1}{{2\pi }}\int\limits_{ - a}^a {J(\eta )\frac{{n\,{z^{n - 1}}}}{{{z^n} - {{(R + i\eta )}^n}}}\,d\eta }.
\end{equation}
With this formula we can compute the flux (in each tape) between the edge $-a$ and an arbitrary point $y<a$ as
\begin{align}
\nonumber
\Phi (y) &= \int\limits_{ - a}^y {{H_x}(t)dt}  = {\mathop{\rm Im}\nolimits} \int\limits_{ - a}^y {H(t)dt}  \\
\nonumber
&=  - \frac{1}{{2\pi }}\int\limits_{ - a}^y {dt} \int\limits_{ - a}^a {J(\eta )\frac{{n\,{{(R + it)}^{n - 1}}}}{{{{(R + it)}^n} - {{(R + i\eta )}^n}}}\,d\eta } \\
\nonumber
&=  - \frac{1}{{2\pi }}\int\limits_{ - a}^a {J(\eta )\,d\eta \,\int\limits_{ - a}^y {\frac{{n\,{{(R + it)}^{n - 1}}}}{{{{(R + it)}^n} - {{(R + i\eta )}^n}}}\,} dt} \\
&=  - \frac{1}{{2\pi }}\int\limits_{ - a}^a {J(\eta )} \,\ln \left| {\frac{{{{(R + iy)}^n} - {{(R + i\eta )}^n}}}{{{{(R + ia)}^n} - {{(R + i\eta )}^n}}}} \right|\,d\eta .\end{align}
From this expression we can derive the integral equation for the current density
\begin{equation}\label{eq:IE_poly}
\rho J(y,t) = \mu d\frac{1}{{2\pi }}\int\limits_{ - a}^a {\dot J(\eta ,t)\ln \left| {{{(R + iy)}^n} - {{(R + i\eta )}^n}} \right|} \,d\eta +C(t).
\end{equation}

\section{Results}\label{sec:results}
In this section we show the ac loss numerical results for various configurations, including a comparison with those obtained with the analytical models~\cite{Mawatari:APL06,Mawatari:APL08}. The plotted ac loss values represent the losses per tape, normalized by the loss value of a single isolated tape carrying the same transport current: it is therefore what we called a geometric factor, representing the change of the ac loss value with respect to that of a single isolated tape, due to the particular geometric configuration under scrutiny.

For our simulations, we considered a tape 12 mm wide with $I_c=330$~A and $n=35$, representative of state-of-the-art YBCO coated conductors. The frequency of the current source was 50~Hz.

For validation purpose, we compared the current density profiles with those obtained by means of the 2-D FEM model developed in~\cite{Brambilla:SST07}, always obtaining an excellent agreement -- an example is shown later in this section.

\begin{figure}[h!]
\centering
\includegraphics[height=6.5 cm]{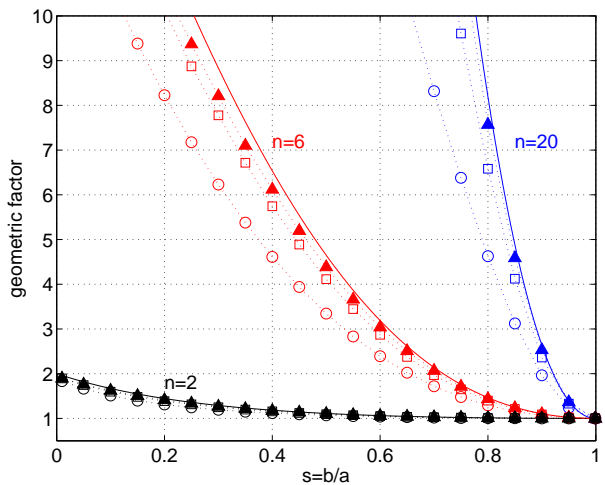}
\caption{Geometric factor indicating the ac losses of a single tape in the radial configuration with unidirectional current with respect to the ac losses of an individual tape carrying the same current. The losses are plotted as a function of the ratio $b/a$ -- see Fig.~\ref{fig:star} for reference. Results are shown for a different number of tapes (2, 6, 20) and for different current values: $I/I_c=0.1$ (triangles), 0.5 (squares) and 0.9 (circles). The continuous line represents the geometric factor calculated by equation (20) in~\cite{Mawatari:APL06}.}
\label{fig:unidir}
\end{figure}

Figure~\ref{fig:unidir} shows the geometric factor of the cable with radially arranged tapes carrying current in the same direction as a function of the distance of the tapes from the center of the cable -- see Fig.~\ref{fig:star} for reference. This is not a convenient geometry from the point of view of ac losses because each tape undergoes the magnetic field generated by the neighbors. At the tape's edge the field contributions sum up, which results in losses that are always higher than those of a single isolated tape. The higher the number of tapes in the cable, the higher the losses.
Plotted with a continuous line is the loss value predicted with the analytical model~\cite{Mawatari:APL06}, which was developed only for the case $I \ll I_c$. It can be noticed that for the case $I=0.1 I_c$ the results of the IE model agree well with the analytical predictions, whereas for higher current values they are significantly different. The difference increases with the number of tapes in the cable.

\begin{figure}[ht]
\centering
\includegraphics[height=6.5 cm]{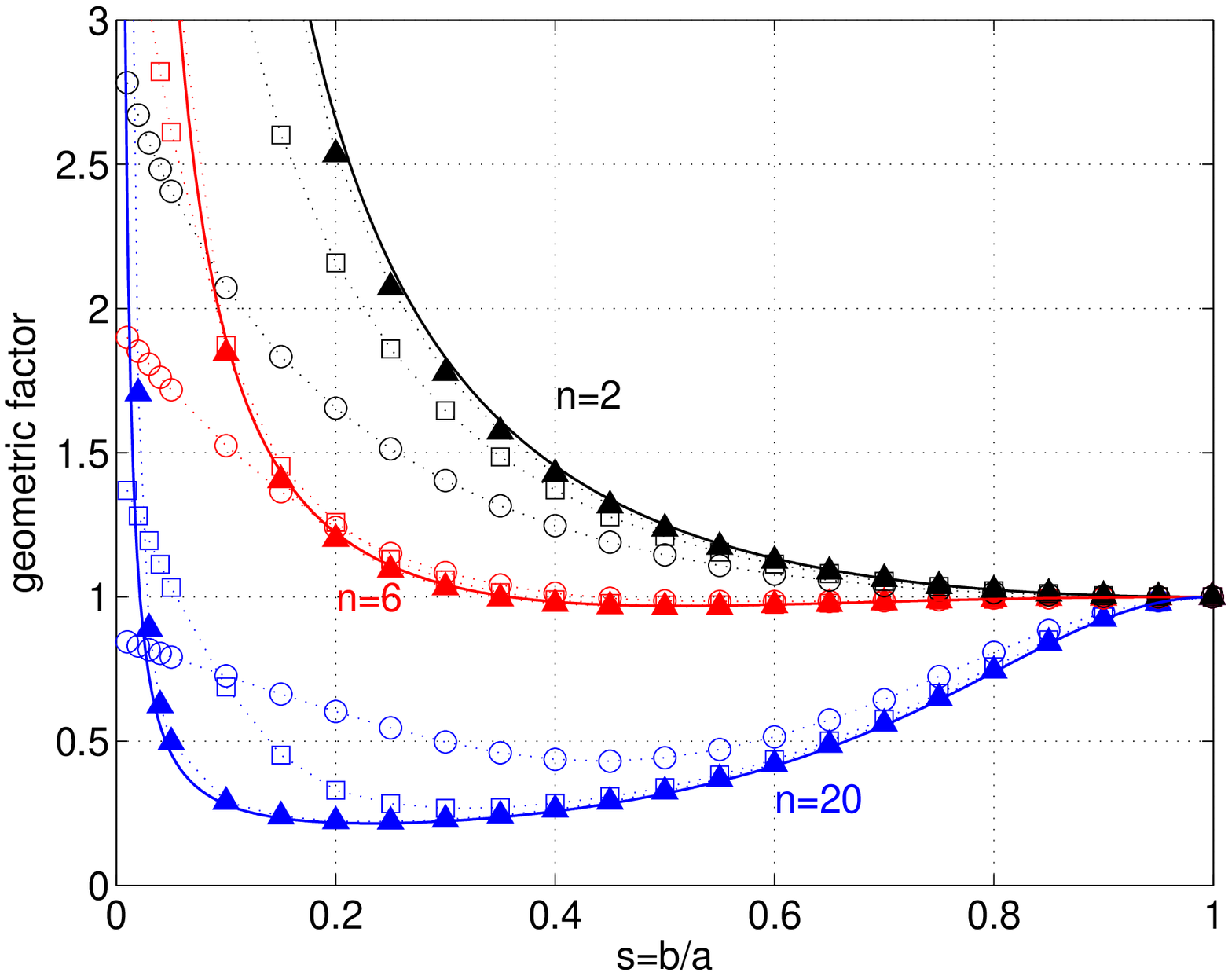}
\caption{Geometric factor indicating the ac losses of a single tape in the radial configuration with bidirectional current with respect to the ac losses of an individual tape carrying the same current. The losses are plotted as a function of the ratio $b/a$ -- see Fig.~\ref{fig:star} for reference. Results are shown for a different number of tapes (2, 6, 20) and for different current values: $I/I_c=0.1$ (triangles), 0.5 (squares) and 0.9 (circles). The continuous line represents the geometric factor calculated by equation (25) in~\cite{Mawatari:APL06}.}
\label{fig:bidir}
\end{figure}

\begin{figure}[h]
\centering
\includegraphics[height=6.5 cm]{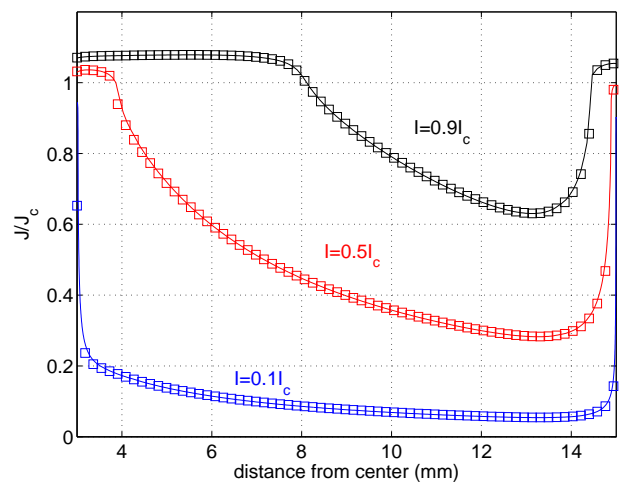}
\caption{Current density profiles in the radially arranged tape configuration for $s=0.2$ ($b=3$ mm, $a=15$ mm) and $n$=20 for different values of the transport current. The profiles are taken at the peak instant of the transport current. The current density is normalized with respect to the critical value $J_c$. For validation purpose, the profiles are calculated with the integral equation model (continuous lines) and with the 2-D FEM model developed in~\cite{Brambilla:SST07} (symbols).}
\label{fig:J_profiles}
\end{figure}

\begin{figure}[ht]
\centering
\includegraphics[height=6.5 cm]{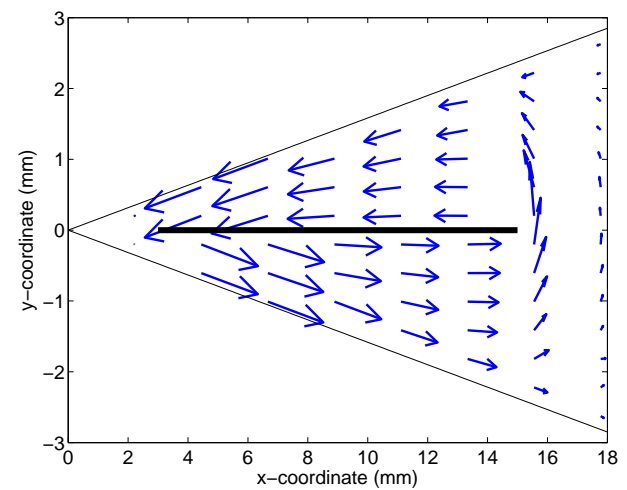}
\caption{Magnetic field generated by a tape in the radial configuration in the case of bidirectional currents: the magnetic field is tangential to the boundaries and the normal component vanishes there. In the case of unidirectional currents, the opposite happens. The represented case corresponds to $n=20$, $b=3$ mm, $a=15$ mm. Different scales are used for the two axes.}
\label{fig:HK_field}
\end{figure}

Figure~\ref{fig:bidir}  shows the same type of results as Fig.~\ref{fig:unidir}, but for bidirectional currents, i.e. adjacent tapes carry the same current but with opposite direction. This is a more advantageous configuration from the point of view of ac losses, because in certain cases the losses are significantly lower than those of a single isolated tape. More specifically this happens for a sufficiently high number of tapes. This is due to the fact that, similarly to the case of bifilar coils~\cite{Clem:PRB08}, the perpendicular component of the magnetic field at the tape's edge is significantly reduced. However, the distance of the tapes from the center also plays an important role and if the tapes are situated very close to the center of the cable ($s \rightarrow 0$) the loss reduction vanishes, especially at high currents.
Also for this configuration we found a good agreement between the IE model and the analytical predictions in the case $I=0.1I_c$ and substantial differences in the case of higher currents.

Figure~\ref{fig:J_profiles} shows the current density profiles for $s=0.2$ ($b=3$ mm, $a=15$ mm) for different values of the transport current. The shape of the profile drastically changes with the current amplitude, and one can clearly see that only for very low currents the Meissner-state approach described in~\cite{Mawatari:APL06} holds. In the figure, results obtained with the 2-D FEM model developed in~\cite{Brambilla:SST07} are also shown (symbols). The overlapping of the profiles computed with the two models is perfect. Due to periodicity, in the 2-D model one needs to simulate only one sector. The condition of uni- or bi-directional currents is obtained by appropriately setting the boundary conditions for the magnetic field (state variable) on the domain's boundary, see also Fig.~\ref{fig:HK_field}. In the case of uni-directional currents, the tangential component of the magnetic field vanishes on the simulated boundary; conversely, in the case of bi-directional currents, the normal component vanishes (as shown in the figure).

\begin{figure}[t]
\centering
\includegraphics[height=6.5 cm]{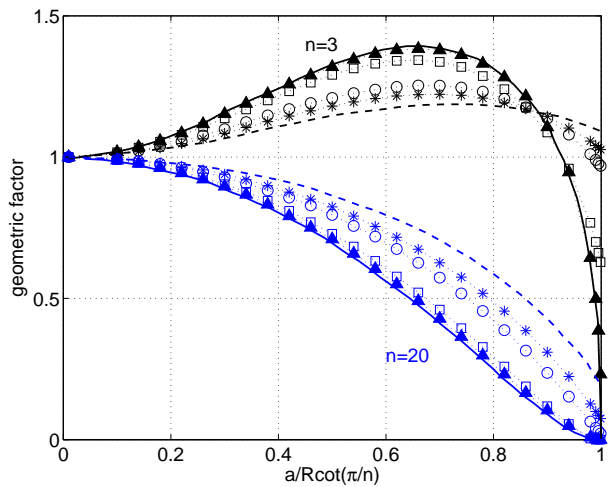}
\caption{Geometric factor indicating the ac losses of a single tape in the polygonal configuration with respect to the ac losses of an individual tape carrying the same current. The losses are plotted as a function of $a/R\cot(\pi/n)$. The continuous and dashed lines indicate the analytical predictions for $I \ll I_c$ and $I=I_c$ from~\cite{Mawatari:APL08}, respectively.} \label{fig:poly}
\end{figure}

Figure~\ref{fig:poly} shows the geometric factor of the losses for a polygonal cable. Losses are plotted as a function of $a/R\cot(\pi/n)$, and in the low current limit they are in excellent agreement with the results of the analytical model in~\cite{Mawatari:APL08}. The agreement with the analytical model is not as perfect in the case $I=I_c$. This is most probably due to the intrinsic difference between the superconductor's characteristics in the two models: with the critical state model (used in the analytical formulas) the tape is filled with current at $I=I_c$ and $J=J_c$ everywhere; on the contrary, with the power-law resistivity (used in the IE model), current density values higher than $J_c$ are allowed and the $J$ is not just constant everywhere.

\section{Conclusions}\label{sec:conclusion}
We derived the integral equations for radially and polygonally arranged thin HTS tapes and solved them by finite elements. The obtained results are in good agreement with existing analytical model, which are valid only for certain values of the transport current. Our integral equation model, on the contrary, can consider arbitrary values of the transport current. Another important advantage of the IE model is that the dependence of the critical current on the lateral position inside the tape or on the local magnetic field can be easily implemented. In addition, since only one tape is simulated in 1-D, the solution takes only few seconds; the IE model can therefore be used to simulate a large number of configurations for design optimization purposes.

Finally, we would like to conclude with a practical remark. Similarly to the cases presented in~\cite{Brambilla:SST09}, the magnetic interaction between the tapes in integral equations~(\ref{eq:IE_unidir}), (\ref{eq:IE_bidir}), (\ref{eq:IE_poly}) is expressed by a term $K(x,t)=\int_{-a}^ak(x-\xi)\dot J(\xi) {\rm d}\xi$, i.e. by the finite space convolution of the time derivative of the sheet current $J(x,t)$ with a kernel $k(x)$ of logarithmic type. This kernel is the only thing that need to be changed to study the different geometries. This means that one needs to build only one file in the finite element program and simply change the selection of the kernel to simulate different configurations.

% use section* for acknowledgement
\section*{Acknowledgment}

The authors would like to thank Dr. Y. Mawatari for sharing useful hints for the implementation of his formulas.

% Can use something like this to put references on a page
% by themselves when using endfloat and the captionsoff option.
\ifCLASSOPTIONcaptionsoff
  \newpage
\fi

% trigger a \newpage just before the given reference
% number - used to balance the columns on the last page
% adjust value as needed - may need to be readjusted if
% the document is modified later
%\IEEEtriggeratref{8}
% The "triggered" command can be changed if desired:
%\IEEEtriggercmd{\enlargethispage{-5in}}

%% references section
%
%% can use a bibliography generated by BibTeX as a .bbl file
%% BibTeX documentation can be easily obtained at:
%% http://www.ctan.org/tex-archive/biblio/bibtex/contrib/doc/
%% The IEEEtran BibTeX style support page is at:
%% http://www.michaelshell.org/tex/ieeetran/bibtex/
%\bibliographystyle{IEEEtran}
%% argument is your BibTeX string definitions and bibliography database(s)
%\bibliography{IE_star_biblio}
%%
%% <OR> manually copy in the resultant .bbl file
%% set second argument of \begin to the number of references
%% (used to reserve space for the reference number labels box)

\end{document}